\documentclass[aps,prl,twocolumn,showpacs,superscriptaddress,10pt]{revtex4-1}

\usepackage{amsmath}
\usepackage{amsfonts}
\usepackage{amssymb}
\usepackage{graphicx}
\usepackage{color}
\usepackage{xcolor}

\begin{document}
\title{Orthogonality catastrophe in dissipative  quantum many body systems}
\author{F. Tonielli}
\affiliation{Institut f\"{u}r Theoretische Physik, Universit\"{a}t zu K\"{o}ln, D-50937 Cologne, Germany}
\author{R. Fazio}
\affiliation{Abdus Salam ICTP, Strada Costiera 11, I-34151 Trieste, Italy}
\affiliation{NEST, Scuola Normale Superiore and Istituto Nanoscienze-CNR, I-56126 Pisa, Italy}

\author{S. Diehl}
\affiliation{Institut f\"{u}r Theoretische Physik, Universit\"{a}t zu K\"{o}ln, D-50937 Cologne, Germany}
\affiliation{Kavli Institute for Theoretical Physics, University of California, Santa Barbara, CA 93106-4030, USA}
\author{J. Marino}
\affiliation{Kavli Institute for Theoretical Physics, University of California, Santa Barbara, CA 93106-4030, USA}
\affiliation{Department of Physics, Harvard University, Cambridge MA 02138, United States}
\affiliation{Department of Quantum Matter Physics, University of Geneva, 1211, Geneve, Switzerland}

\begin{abstract}

We present an analog of the phenomenon of orthogonality catastrophe in  quantum many body systems subject to a local dissipative  impurity.
We show that the fidelity $F(t)$, giving a measure for distance of the time-evolved state from the initial one, displays a universal  scaling form $F(t)\propto t^\theta e^{-\gamma t}$, when the system supports long range correlations, in a fashion reminiscent of traditional instances of orthogonality catastrophe in condensed matter. 
An exponential fall-off at   rate $\gamma$   signals the onset of  environmental decoherence,  {which is critically slowed down by the additional algebraic contribution to the fidelity.}
%
%
~This picture is derived within a second order cumulant expansion suited for Liouvillian dynamics, and substantiated  for the one-dimensional transverse field quantum Ising model subject to a local dephasing jump operator, as well as for XY and XX quantum spin chains, and for the two dimensional Bose gas deep in the superfluid phase with local particle heating.
Our results hint that  local sources of dissipation can be used to inspect   real-time  correlations {and to induce a delay of decoherence} in  open quantum many  body systems.
\end{abstract}


\date{\today}
\maketitle
\emph{Introduction ---}
Anderson's orthogonality catastrophe (OC)~\cite{Anderson} is a paradigm in solid state physics~\cite{Mahan} highlighting  the sensitivity of a  gapless many-body ground state to static and dynamical local perturbations.
An X-ray absorption process creates into an electron gas a core hole which acts as a static potential, provoking a catastrophic  response in the system: 
the ground states of the electron gas, with and without the core-hole potential, are orthogonal -- the overlap between the two scaling as a decaying power law of the system size.
{
Singular features manifest in dynamical properties as well: the Green's function of the core hole has a power law decay at long times, departing from a simple free particle behavior; in frequency domain, close to the threshold energy, the X-ray absorption spectrum vanishes algebraically, signaling the suppression of absorption processes in this energy window. ~\cite{Nozieres69}.}
Orthogonality catastrophe   has been corroborated in a number of systems ranging from Luttinger Liquids~\cite{Gogolin93} to Kondo models~\cite{Yuval70, Yuval70bis} and disordered metals~\cite{Gefen}, and it has recently received novel attention~\cite{Latta11, adilet, 
Munder12, Dora13, Sindona13, Vasseur15, Khemani15, Dora2015, Schmidt17}, thanks to experimental progresses in cold gases, where  local excitations can be created  in a quantum many particle system at ease~\cite{Weitenberg11, Fukuhara13}.

The connection among OC and the return probability,
or Loschmidt echo~\cite{Peres84, Jalabert01, quan, Gamb, heyl, Dora, adilet, dorner, mazzola, Karrasch2010, dpt}, $\mathcal{L}(t)$, is a recent interesting  development in this evergreen problem.
The overlap between the unperturbed ground state of a quantum Ising chain at criticality, $|\psi(0)\rangle$, and the same state evolving in the presence of a defect of strength $\delta g$ along the transverse field direction, $|\psi(t)\rangle$,  exhibits an analogous  algebraic scaling behaviour~\cite{Silva08} to the one discussed above, $\mathcal{L}(t)=\left|\langle\psi(0)|\psi(t)\rangle\right|^2\propto t^{-\theta}$, with $\theta\propto (\delta g)^2$ .
The physical rationale behind the 'catastrophe', stands in the underlying criticality of the many-body system upon which the perturbation is  applied: the diverging characteristic correlation length and times at the critical point, facilitate the spread of the local disturbance across the whole system, making possible the orthogonality among  the initial state and the  evolved one as time increases. 
This set-up can also be extended to non-equilibrium closed environments~\cite{Lupo16}: the system is first sent out of equilibrium by a  quantum quench of a global Hamiltonian parameter, and later subject to the action of a local potential, resulting in a two-times orthogonality catastrophe which may show such novel  features as ageing dynamics~\cite{Schiro14}.

{In this work, we demonstrate that the phenomenon of OC is not only exclusive to \emph{unitary} dynamics, rather it can also occur in a gapless quantum many-body system when a local \emph{noisy} or \emph{dissipative} perturbation  is suddenly switched;  this dissipative analog of the OC is presented through a number of instances ranging from low dimensional quantum spin chains to the Bose-Hubbard model in the superfluid phase.}
%
%
%
%
%
In particular, we show the emergence of a power law scaling in time for the fidelity (a proper analogue of the Loschmidt echo for {generic mixed states}) of a system with  critical,  or, in general, long range correlations, in a fashion reminiscent of the OC in closed  gapless systems.
%
{However, contrary to traditional instances of OC, the additional algebraic contribution to the fidelity determines a critical slow down of decoherence.
%
The paradigm shift presented here for the OC  can be experimentally accessible as localised dissipations can be tailored  in ultracold gases~\cite{ger08, bra12, Ott1, Ott2, Ott3, Ott4, tomita, From}, with the long-run perspective to employ local dissipative channels  to detect gapless modes  in open quantum  many-body systems.
}

\emph{Orthogonality catastrophe from a Lindbladian impurity---}
To illustrate this concept, we consider as a minimal model the one dimensional quantum Ising chain~\cite{Sachdevbook} in a transverse field, $H_0=-\frac{J}{2}\sum_i\left[\hat \sigma^x_i\hat\sigma^x_{i+1}+g\hat\sigma^z_i\right]$, {we} prepare at time $t=0$ the system in its critical ground state ($g=1$) and suddenly switch at subsequent times $t>0$ a spin-dephasing Lindblad operator, $\hat L= \hat \sigma^z_j$, acting on a given site, $j$, of the chain.
Using a Jordan-Wigner transformation~\cite{Sachdevbook}, the critical Ising chain can be mapped into a one-dimensional system of gapless, free fermions with a local dephasing noise, $\hat L\propto \hat n_j$, occurring at rate $\sqrt{\kappa}$,  and proportional to the density $n_j$, of Jordan-Wigner fermions. 
The dynamics of this system is  accordingly ruled by the {Quantum Master Equation {(QME)}}
\begin{equation}\label{eq:lind}
\dot{\rho}(t)=-i[H_0,\rho(t)]+\kappa \mathcal{L}[\rho(t)],
\end{equation}
where $\mathcal{L}[\rho(t)]=\hat{L}\rho(t)\hat{L}^{\dagger}-\frac{1}{2}\{\hat{L}^{\dagger}\hat{L},\rho(t)\}$, and with $\hat{L}=\hat{L}^{\dagger}$, $\hat{L}^2=1$ in this specific case.
%
%
%
The dynamics ruled by the QME with {H}amiltonian, $\hat H_0$, and with a single Hermitian Lindblad operator $\hat L=\hat \sigma^z_j$, is equivalent~\cite{Marinolong2012, Stannigel14} to the stochastic Schr\"{o}dinger evolution governed by 
\begin{equation}
\hat H_{\eta}(t)=\hat H_0+\sqrt{\kappa}\eta(t)\hat L, \label{eq:noisy_Ham}
\end{equation}
where  $\eta(t)$ is a Gaussian white noise and $\hat{L}$ is, for instance, a local  spin perturbation along the transverse field direction, as in the case under study in this work. 
The Lindblad evolution of the density matrix 
$
\hat\rho(t)=e^{t\mathcal{L}}\hat\rho_0=\,\langle \hat U^{ }_{\eta}(t)\hat \rho_0 \hat U_{\eta}^{\dagger}(t)\rangle,
$
can then be recovered averaging over the fluctuations of the white noise, with $\hat U_{\eta}(t)$  the time evolution operator of the time-dependent Schr\"{o}dinger equation at a fixed noise realization $\eta(t)$.
The {H}amiltonian~\eqref{eq:noisy_Ham} renders therefore clearer the connection of our setup to more conventional instances of OC, where algebraic scaling of the Loschmidt echo has been evidenced in  quantum Ising models of the form~\eqref{eq:noisy_Ham} without adding a  noisy character to the local perturbation~\cite{Silva08, Smacchia12}.

\begin{figure}[t!]
\includegraphics[width=9.cm]{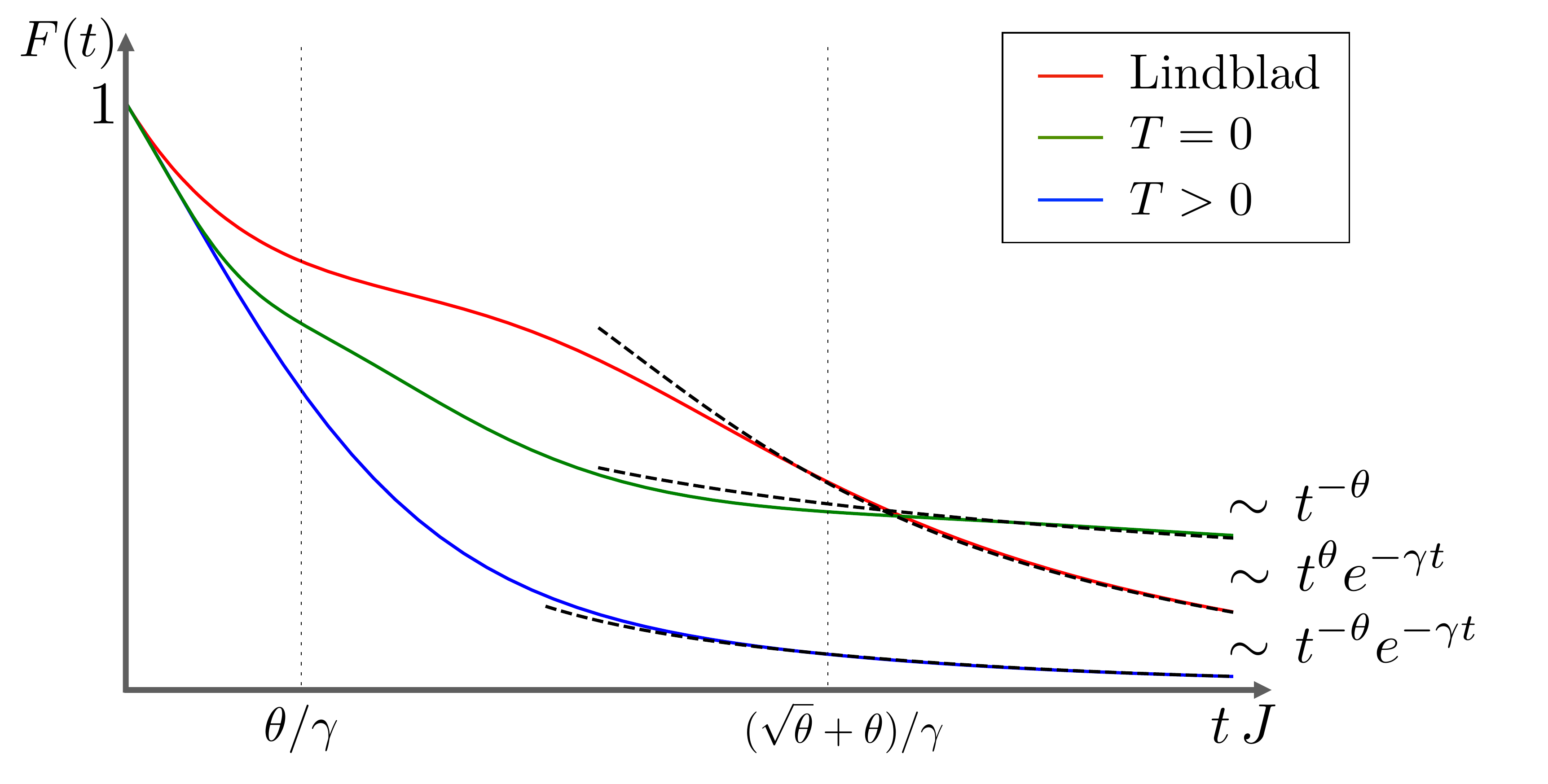}
\caption{(Color online) Comparison between the fidelity $F(t)$ for a quantum Ising chain  with a spin-dephasing impurity (red line) and for the same Ising chain with a local defect on the transverse field, at zero temperature~\cite{Silva08} (green line) and at finite temperature~\cite{Lupo16}  (blue line).  The asymptotic behavior of the three curves is highlighted on the right. {The local lindbladian channel results in a slower decay of the fidelity compared to the other two cases.} 
%
~Close to $t \lesssim(\sqrt{\theta}+\theta)/\gamma$ the fidelity in the dissipative Ising model transits from a concave universal behaviour to the usual convex character typical of isolated systems.} 
\label{fig2}
\end{figure}


However, since the state of the system is mixed at  times $t>0$, we need a generalized expression for the Loschmidt echo in order to investigate the onset of an analogue of OC in the dissipative critical quantum Ising chain. 
A natural choice is represented by the Uhlmann Fidelity~\cite{Uhlmann76, Qinfo}, which reduces to the Loschmidt Echo  when both states 
are pure. 
%
%
If instead only the  initial state  is pure (as in the case under inspection in this work), we  observe that the Uhlmann Fidelity retains a convenient expression
\begin{align}F(t)=\langle\psi(0)|\hat\rho(t)|\psi(0)\rangle=\text{Tr}\left[\hat\rho(0)\hat\rho(t)\right],\end{align}
which is amenable to analytical calculations.
Intuitively, the Loschmidt echo for an open system is equivalent (within Born approximation) to the Uhlmann Fidelity of a given subsystem if the environment remains unaffected during dynamics, since the latter can then be  traced out~\cite{Qinfo} (provided the initial density matrix is a factorised product of the system and environment's density matrices).

At the critical point, the quantum Ising chain reacts  to the presence of the local dephasing channel $\hat{L}$, with  a fidelity   which  decays and scales at long  times as
\begin{align}\label{eq:fid}
F(t) \propto\, t^{+\theta}e^{-\gamma t}, \quad t\gg1/J.
\end{align}
The  power-law character $\propto t^{+\theta}$ recalls the characteristic algebraic response of a gapless quantum system to a local perturbation~\cite{Nozieres69, Silva08, Mahan}, which signals the onset of the phenomenon of orthogonality catastrophe. 
The exponent {$\theta=8/\pi^2 (1-2n)^2 (\kappa/J)^2$}, is, however, \emph{positive}, contrary to  unitary incarnations of OC ($n$ is the local fermion density on the site where the dissipative perturbation is applied, and it is a function of the transverse field, $g$, see for instance~\cite{Sachdevbook}). This brings  the qualitative difference that a new, concave region (see also Fig.~\ref{fig2} and the discussion  in the following section) appears in the {universal} shape of $F(t)$, as a result of the interplay {between $t^{+\theta}$ and}  the exponential decay $\propto e^{-\gamma t}$ with decoherence rate $\gamma=8\kappa n(1-n)$ -- in contrast to the monotonic convex behaviour of the Loschmidt echo in isolated systems.
As in  ordinary instances of orthogonality catastrophe, the power law term is superseded when the many-body environment is away from criticality.
%
%
 %
%

\emph{Cumulant expansion for Lindblad dynamics ---} 
In order to find the long-time  behaviour~\eqref{eq:fid}, we design a second-order cumulant expansion for the fidelity  suited for Lindblad dynamics, which generalises  analogous  methods developed for  the calculation of the Loschmidt echo in isolated systems~\cite{Silva08, Mahan}. 
The key idea is to express $F(t)$ in the super-operator formalism~\cite{Breuerbook}
\begin{align}\label{eq:sup}
F(t)=\,\text{Tr}\left[\hat\rho(0)e^{t\mathcal{L}}\hat\rho(0)\right]\,\equiv\,(\rho_0|e^{t\mathcal{L}}|\rho_0),
\end{align}
where $e^{t\mathcal{L}}$ is the superoperator corresponding to the Lindblad dynamics in \eqref{eq:lind},  acting on the supervector $|\rho_0)$ associated to the initial condition (the ground state of the quantum Ising chain in this specific instance).
Casting $F(t)$ into the form~\eqref{eq:sup}, makes it amenable to a standard perturbative expansion in  interaction picture with respect to the unperturbed (purely {H}amiltonian) Liouvillian $\mathcal{H}_0$, associated to the quantum dynamics of the Ising model. 
Within this representation, we evolve the density matrix, $\hat{\rho}_I(t)\,=\,e^{i\hat{H}_0t}\hat{\rho}(0)e^{-i\hat{H}_0t}$, starting from the critical ground state of the Ising chain,
and we recast the fidelity using $\hat{\rho}_I(t)$ as reference state,
\begin{equation}
F(t)=(\rho_I|T_{\leftarrow}\text{exp}\left\{+\int_0^tds\,\mathcal{L}_I(s)\right\}|\rho_I), 
\end{equation}
which can then be expanded in cumulants ({see Supplemental Material (SM)}),
\begin{align}\label{eq:cumfid}
F(t)=&\,\text{exp}\Big\{+\int_0^tds\,(\mathcal{L}_I(s))^C_0\notag\\
&+\frac{1}{2}\int_0^tds\int_0^tds'\,(T_{\leftarrow}\mathcal{L}_I(s)\mathcal{L}_I(s'))^C_0\,+\,\cdots\Big\}.
\end{align}
In Eq.~\eqref{eq:cumfid}, $T_{\leftarrow}$ is the time ordering operator, $\mathcal{L}_I(s)$ is the Liouvillian perturbation {with its Lindblad operators} evolving under the  {H}amiltonian $\hat H_0$, we used $\hat\rho_I(t)=\hat\rho(0)$ for the initial ground state, and the compact notation $(\cdot)_0\equiv(\rho_0|\cdot|\rho_0)$ has been adopted.
%
%
\begin{figure}[t!]
\includegraphics[width=8cm]{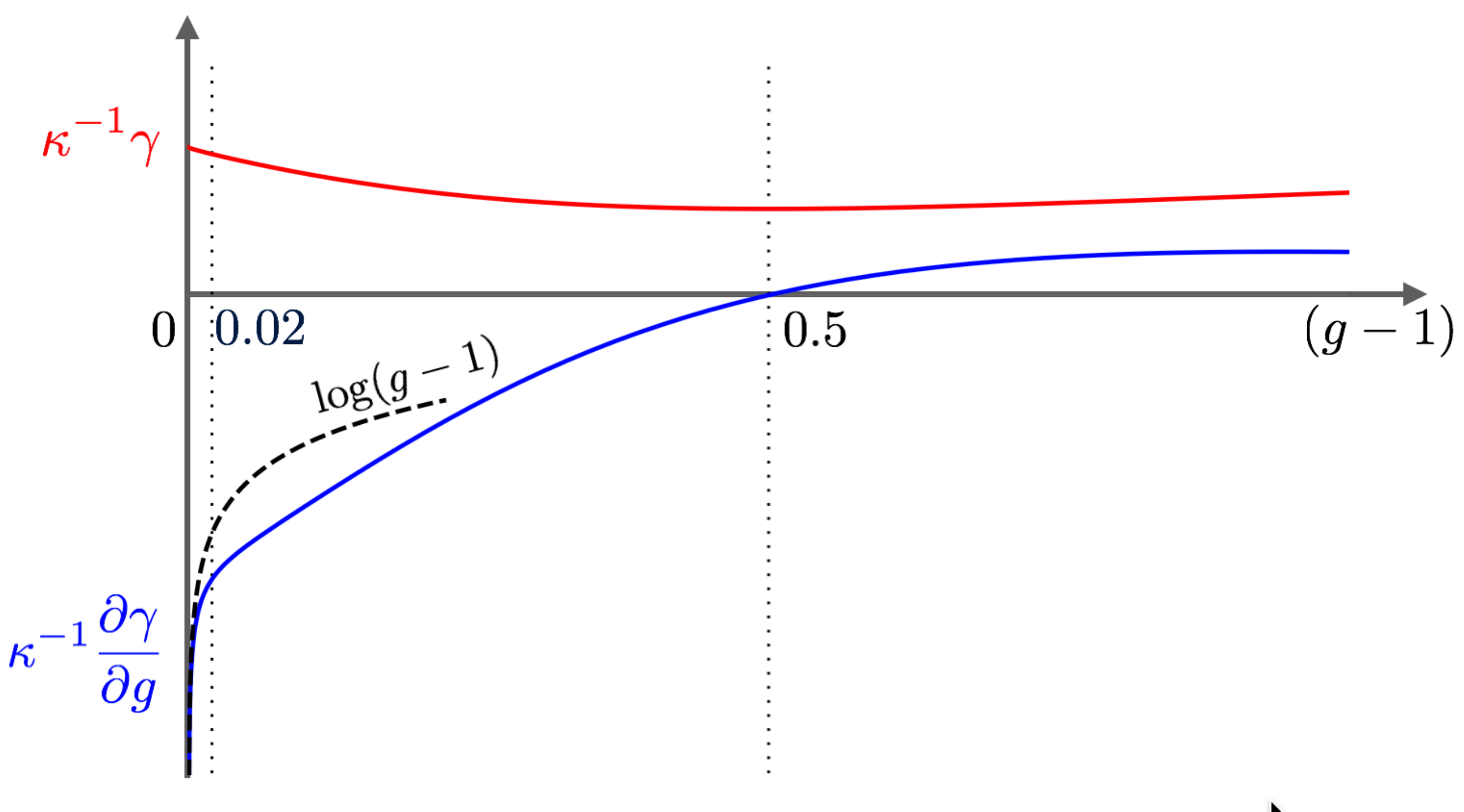}
\caption{(Color online) The rate of exponential decay, $\gamma$, and its derivative as a function of  the gap, $(g-1)$, in the paramagnetic phase of the one dimensional quantum Ising chain subject to local spin dephasing. 
Close to the critical point, $g\to1$, the latter (blue line) exhibits a logarithmic divergence, while the former (red line) is continuous.
}
\label{fig1}
\end{figure}
For a {single, Hermitian} dissipative channel the first two cumulants read
\begin{equation}\label{eq:uno} 
\begin{split}
&(\mathcal{L}_I(s))^C_0\,=\,-2\left(\langle\hat L^2\rangle_0-\langle\hat L\rangle^2_0\right),\\
& (T_{\leftarrow}\mathcal{L}_I(s)\mathcal{L}_I(s'))^C_0\,=\,4\left(|\langle T_{\leftarrow}\hat L(s)\hat L(s')\rangle_0|^2 - \langle \hat L \rangle_0^4\right).
\end{split}
\end{equation}

In order to gain  insight into the first two terms of the  cumulant expansion~\eqref{eq:uno}, we write them  in terms of connected correlation functions of spin operators, 
\begin{align}
\label{eq:firstspin}(\mathcal{L}_I(&s))^C_0\,=\,-2\kappa\big(1- \langle\hat\sigma_j^z\rangle_0^2\big),\\
\label{eq:secondspin}(T_{\leftarrow}&\mathcal{L}_I(s)\mathcal{L}_I(s'))^C_0\,=\ \notag\\
 &=\,8\langle\hat\sigma_j^z\rangle_0^2\,\text{Re}\,G(s-s')+4|G(s-s')|^2,
\end{align}
where  $G(s)=\langle\hat \sigma^z_j(s)\hat \sigma^z_j(0)\rangle - \langle\hat\sigma_j^z\rangle_0^2$.
The first cumulant~\eqref{eq:firstspin} is constant, and when integrated over time yields a term proportional to $t$: this is the exponential decay rate $\gamma$ in Eq.~\eqref{eq:fid}. 
$\gamma$ is continuous close to the critical point $g\to 1$, where it has, however, a  diverging derivative (see also Fig.~\ref{fig1})
\begin{align}
\frac{\partial\gamma}{\partial g}\bigg|_{g\to1}=\frac{8-2\pi}{\pi^2}\kappa \log\left(g-1\right).
\end{align}
This is a first imprint of criticality on the fidelity, although similar features have also been found  in the study of decoherence induced on a two-level system coupled to a one-dimensional quantum spin chain~\cite{Rossini07}.
%
%
%
%
%
%
%

The  second cumulant~\eqref{eq:secondspin} contains, instead, the characteristic features of the OC phenomenon, specifically,
the first contribution to~\eqref{eq:secondspin} diverges logarithmically in $t$ after integration over the variables $s$ and $s'$ (cf. Eq.~\eqref{eq:cumfid}).
Collecting ~\eqref{eq:firstspin} and this leading contribution, we have the following expression for the fidelity ({see SM for details}):
\begin{widetext}
\begin{equation}
F(t)=\text{exp}\Big\{-\gamma t+\kappa^2\int_0^tds\int_0^tds'\,\text{Re}\,G(s-s')+\cdots \Big\} 
=\text{exp}\Big\{-\gamma t+4\kappa^2(1-2n)^2\int_{k,k'}V(k,k')\,\frac{1-\cos(E_k+E_{k'})t}{(E_k+E_{k'})^2}+\cdots \Big\} .
\label{eq:reG}
\end{equation}
\end{widetext}
In Eq.~\eqref{eq:reG}, $V(k,k')=\sin(2\theta_k)\sin(2\theta_{k'})+4\cos^2(\theta_k)\cos^2(\theta_{k'})$ is the same matrix element found in the second order cumulant expansion of~\cite{Silva08}, with $2\theta_k=\operatorname{tg}^{-1}(\sin k/(g-\cos k))$. 
This logarithmic divergence is at the origin of the power law character of~\eqref{eq:fid}, and it can be understood by  power counting (the denominator is $\propto(k+k')^2$, {$V$ is finite}, and integration over momenta $k$,$k'$ is carried out twice).
%
A double time integration over a term $\propto G(s-s')$  appears also in the second cumulant calculation of the Loschmidt echo in an isolated system, causing as well a logarithmic divergence  in time and accordingly the typical algebraic scaling $\sim t^{-\theta'}$~\cite{Nozieres69, Mahan, Silva08, Lupo16}. 
The circumstance that the  same quantity appears in the dissipative setup considered in this work at the same level of cumulant expansion,  confirms the physical intuition that also here critical correlations are the genuine cause of  algebraic scaling.
The same scaling argument shows that the term $\propto |G(s-s')|^2$ from Eq.~\eqref{eq:secondspin} is   subleading with respect to the terms appearing in~\eqref{eq:reG}.
%

%

We finally comment on  the impact of the algebraic scaling $\propto t^{+\theta}$ in $F(t)$ (cf. Eq.~\eqref{eq:fid}). 
The fidelity is always monotonically decreasing, as it should be for a system coupled to a Markovian bath where there cannot be any revival of the information originally present  in the initial state. 
$F(t)$ is apparently increasing for times  $t J\lesssim\theta J/\gamma \propto \kappa/J $, with $\kappa/J\ll 1$,  the small parameter controlling the perturbative cumulant expansion;
however, the algebraic scaling is only valid starting at times of the order $t\sim 1/J$ (as it occurs also in OC for isolated systems~\cite{Silva08}), and therefore {no actual grow occurs.  
Nevertheless, $F(t)$ displays a distinct feature compared to OC phenomena in closed systems: {the presence of a gapless mode provokes the scaling $\propto t^\theta$ and  decoherence is actually slowed down.}
Furthermore, the fidelity at early times is concave, see  Fig.~\ref{fig2} above, and becomes convex at later times. 
The inflection point lies indeed at $t^*J=(\sqrt{\theta}+\theta)/\gamma$, which is $O(1)$ even for $\kappa/J\ll1$. 
This behaviour is general in the sense that it depends only on the long-time  properties of the critical correlations of the model, and it constitutes a novelty of the dissipative scenario.
%


%
%
%

\emph{Other models.} 
We have  tested the emergence of a dissipative analogue of OC in  other systems, ranging from quantum spin chains with conserved local magnetization (XX model) to the two dimensional Bose-Hubbard model with dephasing. 
We focused on the onset of  the scaling term $\propto t^{+\theta}$ contributing to the fidelity, since aspects related to monotonicity and concavity are  based on the generic structure of the perturbative cumulant expansion rather than on specific details of the model at hand. 

The simplest generalisation of the previous setup in one dimension is the XY spin chain~\cite{Lieb61} described by the {H}amiltonian $ 
\hat H_{XY} = -\frac{J}{2}\sum_i\left[\left(\frac{1+\Delta}{2}\right)\hat \sigma^x_i\hat\sigma^x_{i+1}+\left(\frac{1-\Delta}{2}\right)\hat \sigma^y_i\hat\sigma^y_{i+1}+g\hat \sigma^z_i\right]$, with a local dissipative impurity, $\hat{L}=\sigma^z_j$. 
For generic $\Delta\neq0$, analytical results can be obtained from the cumulant expansion of the previous section simply  replacing $\sin k\to\Delta\sin k$.
The latter substitution does not alter the infrared scaling of Eq.~\eqref{eq:reG}, because the quasi-particle energy of the fermions diagonalising $H_{XY}$, has still a linear infrared character as $k\to0$,  $\epsilon_k\sim \Delta |k|$, implying that  the fidelity has an  algebraic scaling contribution also in this model.
When $\Delta=0$,  the {H}amiltonian $H_{XY}$ describes a  $XX$ quantum spin chain~\cite{Sachdevbook}, which conserves the total transverse magnetization ($\hat{M}_z\propto\sum_i\hat{\sigma}^z_i$); the model is therefore equivalent to a system of free fermions in one dimension at  finite density, known to undergo orthogonality catastrophe when coupled to a local potential~\cite{Mahan}.
The dissipative analogue holds as well, the main difference with the Ising case being that the logarithmic divergence in Eq.~\eqref{eq:reG} comes from modes close to the Fermi surface, rather than  from those close to $k=0$. {In passing, this circumstance highlights that criticality is not a necessary condition for the onset of OC: the absence of a gap in the spectrum is sufficient to induce the long-range correlations that cause the algebraic scaling contribution to  the fidelity.}

Finally, we have considered the Bose-Hubbard model~\cite{Sachdevbook}  in  $d$ spatial dimensions, $H_{BH}=-J\sum_{\langle i,j \rangle} \hat{b}^\dag_i \hat{b}_j+\frac{U}{2}\sum_i \hat{n}_i(\hat{n}_i-1)$, {deep in the superfluid phase (where excitations are gapless) and subject to a local heating process described by $\hat{L}_j=\hat{n}_j$}, at rate $\kappa$; despite the model is not at the critical point, the absence of a gap is sufficient to develop long-range correlations which make the model potentially prone to OC.
In the Hartree-Fock-Bogolyubov approximation, the model reduces to a free {H}amiltonian of Bogolyubov quasi-particles; computations follow the perturbative cumulant expansion~\eqref{eq:cumfid} with the additional complication that now  {$\hat L^2\neq \hat {\mathbf{1}}$}, which brings a new term
\begin{equation}
2\text{Re}\bigg[\langle T_{\leftarrow}\hat L^{2}(s)\hat L^{2}(s')\rangle_0-\langle\hat L^{2}\rangle^2_0\bigg],
\end{equation} 
in Eq.~\eqref{eq:secondspin}.
Employing scaling arguments, one can show that the phenomenon of dissipative OC exists only in $d=2$, with a fidelity scaling as $F_{BH}(t)\propto t^{\Theta}e^{-\Gamma t}$, where {$\Gamma\propto\kappa n\left(1+\mathcal{O}(n)\right)$,  $\Theta\propto(\kappa/J)^2(1+4n+\mathcal{O}(n^2))$ and $n$ the density of bosons in the superfluid ground state. In passing, we notice that the interaction strength, $U$, determines the time scales, $t\gg(JUn)^{-1/2}$, for the onset of the scaling form, $F_{BH}(t)$, of the fidelity in the Bose-Hubbard model.
 
\emph{Conclusions and perspectives  ---} 
In summary, we have shown that the {decoherence following the sudden switch of a dissipative impurity on a gapless quantum many-body system is slowed down due to the critical, long range correlations persisting in the system}.
{This phenomenon can be interpreted as another manifestation of the Anderson Orthogonality Catastrophe in the new context of driven-dissipative systems, thanks to the analogy to the stochastic quantum dynamics governed by the Hamiltonian \eqref{eq:noisy_Ham}. In fact, the potential is localized for every realization of the noise, hence transitions it can induce are suppressed in the low-frequency part of the spectrum as result of  conventional Orthogonality Catastrophe physics. The corresponding absorption processes are inhibited in this energy window, and heating is therefore partially slowed, as explicated by the occurrence of a power-law growth $t^{\theta}$ together with the typical exponential decay $e^{-\gamma t}$. }
%
%

{A natural point to address is the transient nature of the phenomenon, i.e., whether dynamics is capable to exit the OC regime at longer times due to heating.
Therefore,} as a future direction, we foresee a calculation of the fidelity for dissipative impurities with non-perturbative or numerical methods, in order to inspect whether its universal shape is a precursor of a pure relaxational regime entirely dominated by decoherence or whether it can persist for asymptotically long times (as it might happen in the context of quantum criticality in driven-dissipative platforms~\cite{DallaTorreDemler2012,Marino2016bis}).

{A further option is represented by the extension of the present study to the case of a non-Markovian impurity or, equivalently, of a non-Markovian noisy transverse field (see  Eq.~\eqref{eq:noisy_Ham} above). In this scenario, a non-monotonic behavior of the fidelity might be realisable due to the backflow of information from the environment to the system; accordingly, an intriguing possibility would be the existence of a time window where an algebraic growth is actually observable, unlike in the present case. }

{There is  currently a research trend which aims at extending traditional topics in statistical mechanics to the domain of dissipative quantum many-body physics, as phase transitions~\cite{Sieberer2016review} or integrability~\cite{banchi,essler,lamac}.
Our work articulates towards this direction; accordingly, a natural next step to substantiate the concept of a dissipative orthogonality catastrophe, would consist in studying the response of driven-open fermionic or bosonic gapless systems~\cite{proseneisert, Moos, Horst} to local disturbances. 
}\\

\emph{Acknowledgements -- } We thank E. Altman, A. Chiocchetta, H.  Fr\"{o}ml, Z. Lenarcic, {A. Rosch,  and} M. Schir\'{o} for interesting discussions on related topics. F.~T.  thanks L. Tolomeo for enlightening mathematical discussions.
S.~D. acknowledges funding by the German Research Foundation (DFG) through the Institutional Strategy of the University of Cologne within the German Excellence Initiative (ZUK 81) and SFB1238 project C04, and by the European Research Council via ERC Grant Agreement n. 647434 (DOQS). J.~M. acknowledges support from the Alexander von Humboldt foundation. JM is supported by the European Union's Horizon 2020 research and innovation
programme under the Marie Sklodowska-Curie grant agreement No 745608 (QUAKE4PRELIMAT). This research was supported in part by the National Science Foundation under Grant No. NSF PHY-1748958.

\bibliography{biblio}

\newpage  
\begin{widetext}

\section{\Large{Supplemental Material}}
\vspace{10pt}

\section{Cumulant expansion of the fidelity for  Liouvillian perturbation}
In the following, we  develop a cumulant expansion for the Loschmidt amplitude following the logic of~[2, 29]. The quantum master equation reads $\dot\rho(t)=-i[H_0,\rho(t)]+\kappa\mathcal{L}[\rho(t)]$; since the unperturbed dynamics is unitary, we can  use the interaction picture with respect to the latter (this is equivalent to perform a change of frame co-rotating  with respect to the unperturbed Hamiltonian):
\begin{align}
\begin{cases}
\rho_{I}(t)\equiv U^{\dagger}(t)\rho(t)U(t),\ \ \ U(t)=e^{-iH_0t}\\
\dot{\rho}_{I}(t)=\kappa\mathcal{L}_I(t)[\rho_{I}(t)],\ \ \,   \ \ \tilde{\mathcal{L}}_I(t)\text{ generated by } L_{I}(t)\equiv U^{\dagger}(t)L\, U(t)
\end{cases}
\end{align}
It is important to observe that this choice of frame does not depend on the assumption that $\rho(t)$ is stationary with respect to $H$; it holds in fact for any initial state.\\
Given an initial state $\rho(0)$, we now wish to compare the state evolved in the presence of the perturbation with the state evolved in the absence of the perturbation, which we label $\rho(t)$ and $\rho_{0}(t)$ respectively for the two cases. It obviously holds $\rho_{0}(0)=\rho(0)$.\\
If we assume the initial state to be pure, the fidelity between the two states takes the simple form  $F(t)=\text{Tr}[\rho_0(t)\cdot\rho(t)]$. Its evaluation becomes then easier in the rotating frame because $\rho_{0}(t)$ does not change in time, i.e., $F(t)=\text{Tr}[\rho(0)\cdot\rho_{I}(t)]$.\\
Furthermore, let us write the time evolution for the density matrix as the time-ordered exponential of an operator acting on a vector. We map density matrices to vectors, $\rho\,\leftrightarrow\, |\rho)$, by choosing a basis for density matrices and associating to each $\rho$ the vector of components on this basis, $|\rho)$. The Liouvillian dynamics is formally represented by:
\begin{subequations}
\begin{align}
&\text{quantum master equation:}\ \begin{cases} |\rho(0))\equiv|\rho_{0})\\ \frac{d}{dt}{|\rho_{I}(t))}=\kappa{\mathcal{L}}_I(t)|\rho_{I}(t))\end{cases}\\
&\text{time evolution operator:}\ \ \ 
|\rho_{I}(t))=\ T_{\leftarrow}e^{\kappa\int_0^tds\,\mathcal{L}_{I}(s)}|\rho_{0})
\end{align}
\end{subequations}

We will not need the explicit construction of $|\rho)$ in the following: it is only necessary to prove eq.~(6) of the main text.\\
We further assume the basis to be orthonormal with respect to the Hermitian product $\langle A,B\rangle = \text{Tr}[A^{\dagger}B]$. Under this assumption, one can express the fidelity as 
\begin{align}
F(t)=\,&\text{Tr}[\rho(0)\cdot\rho_{I}(t)]=(\rho_{0}|\rho_{I}(t)),
\end{align}
where the latter is the usual Hermitian product in Hilbert spaces.\\
Let us combine the two results and get:
\begin{align}
F(t)=\,&(\rho_{0}|\rho(t))=(\rho_{0}|T_{\leftarrow}e^{\kappa\int_0^tds\mathcal{L}_{I}(s)}|\rho_{0})=\notag\\
=\,&\exp\left\{\kappa\,\int_0^tds (\mathcal{L}_{I}(s))_0+\frac{\kappa^2}{2}\int_0^t\int_0^t dsds'(T_{\leftarrow}\mathcal{L}_{I}(s)\mathcal{L}_{I}(s'))^C_0+O(\kappa^3)\right\},
\end{align}
where $(\cdots)_{0}=(\rho_{0}|\cdots|\rho_{0})$, and $(\cdots)^{C}_{0}$ denotes the usual connected correlation function:
\begin{align}
(T_{\leftarrow}\mathcal{L}_{I}(s)\mathcal{L}_{I}(s'))^C_0&=(T_{\leftarrow}\mathcal{L}_{I}(s)\mathcal{L}_{I}(s'))_0-(\mathcal{L}_{I}(s))_0(\mathcal{L}_{I}(s'))_0.
\end{align}
Ordinary traces can be immediately restored in the above expression by transforming vectors again in density matrices, and by acting with the Liouvillian operator on all the vectors on its right as the Liouvillian would act on density matrices. 
Some general observations are in order:
\begin{itemize}
\item The Lindbladian part of the Liouvillian comes from second order perturbation theory applied to the Hamiltonian of a system coupled to a bath; it follows that the perturbative expansion in $\delta\mathcal L$ at a given order, $n$, corresponds to that of an expansion in $\delta H$ with order  $2n$. As a consequence, pre-factors of the expansion are never imaginary,  and proportional to $i^{2n}=(-1)^n$; this is consistent with the reality of the fidelity, that is, of the square modulus of the Loschmidt amplitude.
\item The first cumulant yields a real exponential of the form $e^{-\gamma t}$. Let us prove that $\gamma\geq0$, a necessary condition for the expansion to be consistent with the general property $F(\rho,\sigma)\leq 1$. $\mathcal{L}$ can contain Hamiltonian perturbations $V$ and Lindbladian perturbations with Lindblad operators $L$. The statement holds for each:
\begin{subequations}
\label{eq:firstcumulant-hamiltonian+lindblad}
\begin{align}
(\rho_0|\,\mathcal{H}_{I}(s)[\cdot]\,|\rho_0)=&\,(\rho_0|\,-i[V(s),\cdot]\,|\rho_0)=\notag\\
=&-i\text{Tr}\left[\rho_0[V(s),\rho_0]\,\right]=\notag\\
=&-i\text{Tr}\left[\rho_0V(s)\rho_0-\rho_0^2V(s)\right]=0; \label{uno}\\
(\rho_0|\,\mathcal{D}_{I}(s)[\cdot]\,|\rho_0)=&\,\text{Tr}\left[2\rho_0L(s)\rho_0L^{\dagger}(s)-2\rho_0L^{\dagger}(s)L(s)\rho_0\right]=\notag\\=&\,2\left(\langle\psi|L(s)|\psi\rangle\langle\psi|L^{\dagger}(s)|\psi\rangle-\langle\psi|L^{\dagger}(s)L(s)|\psi\rangle\right)=\notag\\
=&-2\left\langle\big(L(s)-\langle L(s)\rangle_{0}\big)^{\dagger}\big(L(s)-\langle L(s)\rangle_{0}\big)\right\rangle
\end{align}
\end{subequations}
These formulas are exact and do not rely on the assumption of stationarity of $\rho_{0}$. We see that, at all times, $(\mathcal{L}_{I}(s))_0\geq0$, consistent with an exponential decay. If the state is stationary with respect to the unperturbed Hamiltonian dynamics, equations simplify considerably and time dependence drops out, yielding for the first cumulant the simple result 
\begin{align}
-2\kappa \left\langle\big(L-\langle L\rangle_{0}\big)^{\dagger}\big(L-\langle L\rangle_{0}\big)\right\rangle\cdot t
\end{align}
Eq.~(6a) can be surprising at a first sight, because it tells us that Hamiltonian perturbations always have a vanishing first cumulant; this is not the case e.g. in [29]. However, such first order terms are always imaginary corrections to the exponent of the Loschmidt amplitude, and they drop out here because we are computing the \emph{square modulus} of the latter.\\
Eq.~(6b) clearly indicates that the first Lindblad cumulant is never positive. It can be negative or zero: the latter case is actually trivial. In fact, from eq.~(6b) a vanishing first cumulant would imply that the variance of $L$ on the initial state is zero. This is possible if and only if $L|\psi_0\rangle = 0$, in which case there is no nontrivial dynamics induced by $L$ because the initial state is already a dark state for the dissipative evolution.
\item As explicated by eq.~(7), the first cumulant gives a trivial contribution to the exponent of the fidelity and determines the leading contribution to the dephasing rate. Any nontrivial feature, including the algebraic behavior, can only come from higher cumulants (the second, in particular); these correspond to fourth or higher order processes in Hamiltonian perturbation theory. This is in slight contrast with the traditional instance of the OC; nevertheless, the effective mechanism behind the algebraic factor is still the scattering of a single particle-hole pair against the impurity, as in the traditional OC.  

\end{itemize}

\vspace{10pt}
\section{Ising model with local dephasing}
In the following, we shall compute the first and the second cumulant in the  case of the critical quantum Ising model perturbed by local heating, discussed in the main text. We use the following notation: $\sigma^{z}_{0}(s)$ are spin operators $\sigma^{z}$ localized at site $j=0$ and in the Heisenberg picture with respect to the Ising Hamiltonian; $f_{0}^{(\dagger)}(s)$ are the annihilation (creation) operators for the Jordan-Wigner fermions in the Heisenberg picture.
\vspace{5pt}
\subsection{First cumulant:}
We  need eq. \eqref{eq:firstcumulant-hamiltonian+lindblad} and the explicit form of the Lindblad operator:
\begin{subequations}
\begin{align}
&L=\sigma^z_0=1-2f^{\dagger}_0f^{ }_0& &\Rightarrow& &(\mathcal{L})_0=-2\left(\langle (\sigma^z_0)^2\rangle_0 - \langle \sigma^z_0 \rangle_0^2\right)=-2\left(1 - \langle \sigma^z_0 \rangle_0^2\right)=\\
& & & & &\ \ \ =-2\left(1-1+4\langle f^{\dagger}_0f^{ }_0\rangle_0-4\langle f^{\dagger}_0f^{ }_0\rangle_0^2\right)= 
 -8 \langle f^{\dagger}_0f^{ }_0\rangle_0\left(1-\langle f^{\dagger}_0f^{ }_0\rangle_0\right)=\notag\\
& & & & &\ \ \ =-8n(1-n).
\end{align}
\end{subequations}

In the thermodynamic limit,
\begin{align}\label{eq:density-Ising}
n=\int_k\langle f^{\dagger}_k f^{ }_k \rangle_0=\int_{-\pi}^{\pi}\frac{dk}{2\pi}\,\frac{g-\cos k}{\sqrt{1+g^2-2g\cos(k)}}
\end{align}
and the integral function of $g$ is continuous. In particular,  close to $g=1$, we find $n=\frac{2}{\pi}$.\\

The expression of $n$ as a function of $\delta=g-1$ reads
\begin{align}
n(\delta)=\frac{2}{\pi}\,\text{K}\left(-\frac{4(1+\delta^2)}{\delta}\right)\,-\frac{i}{\pi}\Bigg(&\frac{2+\delta}{1+\delta}\,\text{E}\left(\frac{\delta^2}{(2+\delta)^2}\right)-\frac{\delta}{1+\delta}\,\text{E}\left(\frac{(2+\delta)^2}{\delta^2}\right)-\frac{4}{2+\delta}\,\text{K}\left(\frac{\delta^2}{(2+\delta)^2}\right)\Bigg),
\end{align}
where $K(x)$ and $E(x)$ are complete elliptic integrals of the first and second kind respectively.\vspace{5pt}
There is a logarithmic singularity of the first derivative of n at the critical point:
\begin{align}
\frac{\partial n}{\partial g}&=\int_{-\pi}^{\pi}\frac{dk}{2\pi}\,\left[\frac{1}{\sqrt{1+g^2-2g\cos(k)}}-\frac{g-\cos(k)}{(1+g^2-2g\cos(k))^{3/2}}\right]\underset{g-1\equiv\delta}{=}\notag\\
&=\int_{-\pi}^{\pi}\frac{dk}{2\pi}\,\left[\frac{1}{\sqrt{\delta^2+4(1+\delta)\sin^2(k/2)}}-\frac{\delta+2\sin^2(k/2)}{(\delta^2+4(1+\delta)\sin^2(k/2))^{3/2}}\right]
\end{align}
In particular, we have 
\begin{align}
\frac{\partial n}{\partial g}\bigg|_{g\to1}\sim\frac{2}{\pi\delta}\,\text{K}\left(-\frac{4(1+\delta^2)}{\delta}\right)\sim-\frac{1}{\pi}\log(\delta)+\mathcal{O}(1)
\end{align}
A similar feature has been pointed out in Ref.~[46] for the rate of the Gaussian decay of the Loschmidt amplitude.\\
The decay rate is therefore also singular: 
\begin{align}
\gamma = 2\kappa n(1-n)\,\Rightarrow \frac{\partial \gamma}{\partial g}\bigg|_{g\to1} = 2\kappa(1-2n)\,\frac{\partial n}{\partial g}\bigg|_{g\to1} \sim\, \frac{2(4-\pi)}{\pi^{2}}\kappa\log(\delta)+\mathcal{O}(1),
\end{align}
as reported in eq.~(11) in the main text.\\
 \vspace{5pt}
\subsection{Second cumulant:}
We need to expand $(T_{\leftarrow}\tilde{\mathcal{L}}_s\tilde{\mathcal{L}}_{s'})^C_0$ and then to integrate over time twice. We assume without loss of generality $s>s'$:
\begin{align}
(\tilde{\mathcal{L}}_s\tilde{\mathcal{L}}_{s'})_0&=\text{Tr}\left[\tilde{\mathcal{L}}_s^{\dagger}[\rho_0]\tilde{\mathcal{L}}_{s'}^{ }[\rho_0]\right]=\notag\\
&=\text{Tr}\left[\left(2\tilde L|\psi\rangle\langle\psi|\tilde L^{\dagger}-\tilde L^{\dagger}\tilde L|\psi\rangle\langle\psi|-|\psi\rangle\langle\psi|\tilde L^{\dagger}\tilde L\right)_s\cdot\left(2\tilde L^{\dagger}|\psi\rangle\langle\psi|\tilde L-\tilde L^{\dagger}\tilde L|\psi\rangle\langle\psi|-|\psi\rangle\langle\psi|\tilde L^{\dagger}\tilde L\right)_{s'}\right]=\notag\\
&=4\langle\tilde L^{\dagger}(s)\tilde L^{\dagger}(s')\rangle\langle \tilde L(s')\tilde L(s)\rangle+2\langle L^{\dagger}L\rangle^2\notag\\
&-4\text{Re}\left[\langle \tilde L^{\dagger}(s)\tilde L^{\dagger}(s')\tilde L(s')\rangle\langle\tilde L(s)\rangle\right]-4\text{Re}\left[\langle \tilde L^{\dagger}(s)\tilde L(s)\tilde L^{\dagger}(s')\rangle\langle\tilde L(s')\rangle\right]\notag\\
&+2\text{Re}\left[\langle\tilde L^{\dagger}(s)\tilde L(s)\tilde L^{\dagger}(s')\tilde L(s')\rangle\right]. \label{eq:secondcumulant-lindblad}
\end{align}
Contractions of operators belonging to the same Liouvillian are canceled by subtracting $(\tilde{\mathcal{L}}_s)_0(\tilde{\mathcal{L}}_{s'})_0$ because the cumulant only keeps the connected part; the book-keeping rule is that contraction schemes involving only operators with the same time argument have to be ignored. The above expression contains correlators with  2, 4, 6 or 8 fields in the case of $L\sim \sigma^z_0$.\\
For the sake of computing these correlators using Wick's theorem, let us write the relevant Green's functions matrix:
\begin{align}
G(s-s')\,&\equiv\,\left \langle T_{\leftarrow}\begin{pmatrix}f^{ }_0(s)f^{\dagger}_0(s') & f^{\dagger}_0(s)f^{\dagger}_0(s') \\ f^{ }_0(s)f^{ }_0(s') &f^{\dagger}_0(s)f^{ }_0(s')\end{pmatrix}\right\rangle\,=\,\int_k \left\langle T_{\leftarrow}\begin{pmatrix}f^{ }_k(s)f^{\dagger}_k(s') & f^{\dagger}_{-k}(s)f^{\dagger}_k(s') \\ f^{ }_k(s)f^{ }_{-k}(s') &f^{\dagger}_{-k}(s)f^{ }_{-k}(s')\end{pmatrix} \right\rangle = \notag\\
&=\,\frac{1}{2}\int_k e^{-iE_k |s-s'|}\begin{pmatrix} \epsilon_k/E_k+\textsl{sgn}(s-s') & +i\Delta_k/E_k \\ -i\Delta_k/E_k & -\epsilon_k/E_k+\textsl{sgn}(s-s') \end{pmatrix}\,\equiv\,\frac{1}{2}\tilde G(s-s')
\end{align}
where
\begin{align}
\begin{cases}
\epsilon_k=g-\cos(k),\ \ \ \Delta_k=\sin(k),\ \ \ E_k=\sqrt{\epsilon_k^2+\Delta^2_k}\\
\epsilon_k/E_k=\cos(2\theta_k),\ \ \ \Delta_k/E_k=\sin(2\theta_k)
\end{cases}
\end{align}
and $\textsl{sgn}(0)=1$.\\

In the present case, calculations are considerably simplified by $\hat{L}=\hat{L}^{\dagger}$ and $\sigma^z(s)\cdot\sigma^z(s)=1$: the second and the fifth term of  eq.~\eqref{eq:secondcumulant-lindblad} become constants, and the third and fourth become proportional to the square of the expectation value of $\sigma^z(s)$. Therefore, 3 and 4 point correlation functions reduce to constants. Furthermore, eq.~\eqref{eq:secondcumulant-lindblad} simplifies even more after subtraction of the connected component. Expectation values are kept only if they involve at least two operators with different time arguments: the third and fourth terms disappear and only part of the first has to be kept.\\

Let us now compute the first term explicitly:
\begin{align}
&4\langle\tilde L^{\dagger}(s)\tilde L^{\dagger}(s')\rangle\langle \tilde L(s')\tilde L(s)\rangle\notag=\\
=&4\left|\left\langle\sigma^z_0(s)\sigma^z_0(s')\right\rangle\right|^2=\notag\\
=&4\left|\left\langle(1-2f^{\dagger}_0(s)f_0(s))(1-2f^{\dagger}_0(s')f_0(s'))\right\rangle\right|^2=\notag\\
=&4\left|1-4n+4n^2+4\left\langle f^{\dagger}_0(s)f_0(s)f^{\dagger}_0(s')f_0(s')\right\rangle_C\right|^2=\notag\\
=&4(1-2n)^4+32(1-2n)^2\text{Re}\left\langle f^{\dagger}_0(s)f_0(s)f^{\dagger}_0(s')f_0(s')\right\rangle_C+64\left|\left\langle f^{\dagger}_0(s)f_0(s)f^{\dagger}_0(s')f_0(s')\right\rangle_C\right|^2. \label{eq:firstterm-secondcumulant-sigmaz}
\end{align}
The first term in the last expression is a disconnected part. The second and third look similar but have a different physical interpretation. The connected correlation function is the same cumulant appearing in~[29]; in momentum space:
\begin{subequations}
\begin{align}
32&(1-2n)^2\,\frac{1}{2}\int_0^t ds \int_0^t ds'\,\text{Re}\left\langle f^{\dagger}_0(s)f_0(s)f^{\dagger}_0(s')f_0(s')\right\rangle_C\,=\notag\\
&=\,8(1-2n)^2\int_{k,k'}\frac{1-\cos(E_k+E_{k'})t}{(E_k+E_{k'})^2}\cdot\left(1-\frac{\epsilon_k\epsilon_{k'}}{E_kE_{k'}}\right),\\
\notag\\
64&\,\frac{1}{2}\int_0^t ds \int_0^t ds'\int_{s,s'}\left|\left\langle f^{\dagger}_0(s)f_0(s)f^{\dagger}_0(s')f_0(s')\right\rangle_C\right|^2\,=\notag\\
&=64\int_0^tdD\,(t-D)\int_{k_1\cdots k_4}\frac{1}{2^4}\,e^{-i(E_{k_1}+E_{k_2}-E_{k_3}-E_{k_4})D}\,\notag\\
&\ \ \ \ \ \ \ \ \cdot \left(1-\frac{\epsilon_{k_1}\epsilon_{k_2}+\Delta_{k_1}\Delta_{k_2}}{E_{k_1}E_{k_2}}+\textsl{sgn}(D)\left(\frac{\epsilon_{k_1}}{E_{k_1}}-\frac{\epsilon_{k_2}}{E_{k_2}}\right)\right)\notag\\
&\ \ \ \ \ \ \ \ \cdot\left(1-\frac{\epsilon_{k_3}\epsilon_{k_4}+\Delta_{k_3}\Delta_{k_4}}{E_{k_3}E_{k_4}}+\textsl{sgn}(D)\left(\frac{\epsilon_{k_3}}{E_{k_3}}-\frac{\epsilon_{k_4}}{E_{k_4}}\right)\right)\notag=\\
&=4\int_{k_1\cdots k_4}\frac{1-\cos(E_{k_1}+E_{k_2}-E_{k_3}-E_{k_4})t}{(E_{k_1}+E_{k_2}-E_{k_3}-E_{k_4})^2}\cdot\left(1-\frac{\epsilon_{k_1}\epsilon_{k_2}}{E_{k_1}E_{k_2}}\right)\cdot\left(1-\frac{\epsilon_{k_3}\epsilon_{k_4}}{E_{k_3}E_{k_4}}\right).\label{eq:integral2}
\end{align}
\end{subequations}
The two terms clearly have a different physical interpretation. The first is the real time amplitude for scattering of a single particle-hole pair against the impurity, whereas the second describes a process of scattering of two pairs. A naive conclusion would be that the first process is parametrically suppressed at fourth order in perturbation theory: the phase space for a single scattering is in fact smaller than the corresponding one for a double process. It turns out in this case that the uniform  background crucially \emph{enhances the phase space volume}, equalizing the magnitude of the two terms. The presence of a uniform background turns out to be always crucial for the decoherence catastrophe.

As a side technical comment, we mention the manipulations we used to obtain these expressions: parity $k_i\to-k_i$ makes off-diagonal terms vanish; reflections $k_1\leftrightarrow k_2$ and $k_3 \leftrightarrow k_4$ cancel terms proportional to $\textsl{sgn}(D)$ inside each bracket; the product of the curly brackets is even under $\{k_1,k_2\}\leftrightarrow \{k_3,k_4\}$, canceling the (odd) imaginary contribution that would have come from the integration over time of the exponential.

Finally, we can study the two integrals eq.~(18a) and eq.~(18b) separately.
\begin{itemize}
\item The denominator in (18a) is always bounded except for $k,k'\simeq 0$: by Riemann-Lebesgue's lemma, the integral over a region that excludes $k,k'\simeq 0$ converges to a finite value for $t\to\infty$. We can therefore extrapolate the singularity by restricting the integration region to $-\Lambda\leq k,k'\leq\Lambda$, $\Lambda\ll1$, and by approximating the integrand. The integral becomes:
\begin{align}
 8(1-2n)^{2}\left[4\int_{0}^{\Lambda}\frac{dk}{2\pi}\int_{0}^{\Lambda}\frac{dk'}{2\pi}\,\frac{1-\cos(k+k')t}{(k+k')^{2}}+\mathcal{O}(1)\right]\,\simeq\,\frac{8}{\pi^{2}}(1-2n)^{2}\log(t)+\mathcal{O}(1),
\end{align}
as naive power counting of the IR divergence suggested. Eq.~(18a) therefore diverges logarithmically and reproduces OC physics as expected.

\item The integral (18b) is much trickier: the denominator vanishes on a complicated manifold that extends throughout the whole Brillouin Zone, not only close to $k_{i}\simeq 0$. A convenient strategy is to compute the time derivative of eq.~(18b) rather than eq.~(18b) itself because the former is bounded for large $t$, whereas the latter is not.\\
We take the time derivative and obtain
\begin{align}
\mathcal{I}\equiv4\int_{k_1\cdots k_4}\frac{\sin(E_{k_1}+E_{k_2}-E_{k_3}-E_{k_4})t}{(E_{k_1}+E_{k_2}-E_{k_3}-E_{k_4})}\cdot\left(1-\frac{\epsilon_{k_1}\epsilon_{k_2}}{E_{k_1}E_{k_2}}\right)\cdot\left(1-\frac{\epsilon_{k_3}\epsilon_{k_4}}{E_{k_3}E_{k_4}}\right)
\end{align}
If the integral converges to a finite constant for $t\to\infty$, the leading divergence of eq.~(18b) is linear in $t$ and the prefactor is the constant times $\kappa^{2}$. This is a rigorous statement. If the subleading correction is of order $t^{-1}$, we expect this to give rise to a logarithmic divergence when integrated over time, albeit this is not a rigorous statement: oscillating terms, e.g. $t^{-1}\sin(t)$, converge to finite constants when integrated over $t$. We fit the tail with constant plus subleading corrections, including oscillatory ones. This procedure identifies an \emph{approximate value} to the contribution to $\theta$ coming from this integral: this will be enough for an order of magnitude estimate, which is all we need. We point out that the same strategy would give the correct value if applied to the previous case.\\
The numerical result for $\mathcal{I}$ can be obtained the following way. We first change variables $k_{i}\to E_{k_{i}}\equiv 2x_{i}$:
\begin{align}
&x = E_{k}/2 = \sqrt{(1-\cos k)/2}\ \ \Rightarrow\ \ dk = \frac{2x}{\sqrt{1-x^{2}}}\,dx \notag\\
\mathcal{I} = 2\left(\frac{2}{\pi}\right)^{4}\int_{0}^{1}&dx_{1}\cdots dx_{4}\ \frac{\sin[(x_{1}+x_{2}-x_{3}-x_{4})2t]}{(x_{1}+x_{2}-x_{3}-x_{4})}\,(1-x_{1}x_{2})(1-x_{3}x_{4})\,\prod_{i}\left(\frac{x_{i}}{\sqrt{1-x_{i}^{2}}}\right)
\end{align}

We then integrate over the differences $x_{1}-x_{2}$ and $x_{3}-x_{4}$ analytically:
\begin{align}
&u\equiv x_{1}+x_{2},\ \ \ r\equiv x_{1}-x_{2}\ \ \ \Rightarrow\ \ \ 0\leq u \leq 2,\ \ \ 0\leq r\leq \text{min}(u,2-u)\notag\\
&(1-x_{1}x_{2})\frac{x_{1}x_{2}}{\sqrt{(1-x_{1}^{2})(1-x_{2}^{2})}} = (1-(u^{2}-r^{2})/4)\frac{(u^{2}-r^{2})/4}{\sqrt{1-(u^{2}+r^{2})/2 + (u^{2}-r^{2})^{2}/16}}\notag\\
\varphi(u)\equiv&\int_{0}^{\text{min}(u,2-u)}\frac{(1-(u^{2}-r^{2})/4)(u^{2}-r^{2})/4}{\sqrt{1-(u^{2}+r^{2})/2 + (u^{2}-r^{2})^{2}/16}}\,dr
\end{align}
and in the end $\varphi(u)$ can be expressed as an involved combination of incomplete elliptic integrals, which we do not report.\\
Finally, we can perform a numerical integration over $u$ and $v\equiv x_{3}+x_{4}$, for times up to $t\sim25$ and time step $\Delta t=0.5$:
\begin{align}
\mathcal{I}(t) = \frac{1}{2}\left(\frac{2}{\pi}\right)^{4}\int_{0}^{2}du\int_{0}^{2}dv\,\frac{\sin[2t(u-v)]}{u-v}\,\varphi(u)\varphi(v).
\end{align}
The integral converges to a finite value; we fit the tail ($t>12.5$) with the function \begin{align}
f(t)=c_{0}+c_{1}t^{-1}+c_{1c}\cos(t)t^{-1}+c_{1s}\sin(t)t^{-1}+c_{2}\,t^{-2},
\end{align}
the interesting coefficients being the first two. The result is depicted in Fig.~1.
 \begin{figure}[t!]
\includegraphics[width=11.cm]{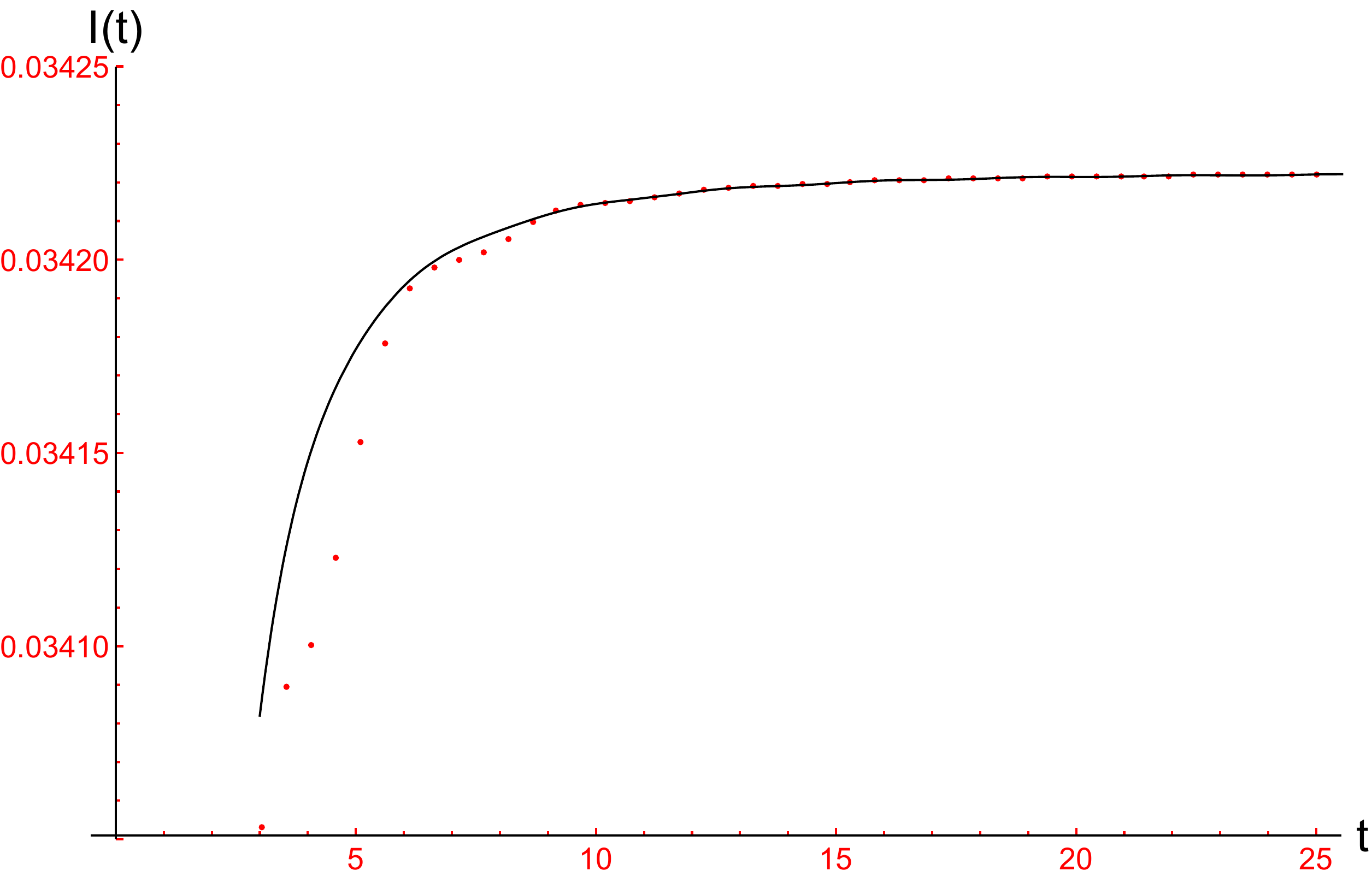}
\caption{(Color online) Numerical evaluation of $\mathcal{I}(t)$  (red dots) for times up to $t=25$ and with time step $\Delta t=0.5$. Function (24) fitted for $t>12.5$ (black line).  }
\end{figure}

The constant term is $c_{0}=0.034$, hence eq.~(18b) scales at long times as $+0.034\,\kappa^{2}\,t$. This is a second order contribution to the decay rate and sets a bound for the validity of the cumulant expansion: from $|\gamma^{(2)}|\lesssim |\gamma^{(1)}|$, it follows
\begin{align}
&0.034\,\kappa^{2}\,\lesssim\,\frac{4(\pi-2)}{\pi^{2}}\,\kappa\ \ \ \ \ \ \Rightarrow\ \ \ \ \ \  \kappa\,\lesssim\,13.5.
\end{align}
The coefficient of $t^{-1}$ is an estimate of the order of magnitude of the prefactor of $\log(t)$; however, $c_{1}=+7\cdot 10^{-5}$ is three orders of magnitude smaller than the analogous coefficient in eq.~(19), and is therefore negligible in any case. We point out that this is the concrete realization of the phase space suppression mechanism mentioned above: the ``OC-like'' contribution coming from the four-particle scattering is much smaller than the corresponding contribution coming from the two-particle scattering.

The coefficients $c_{1\,c,s}$ are two orders of magnitude smaller than $c_{1}$; correspondingly, fitted values for $c_{0,1,2}$ do not change if the oscillatory functions are excluded from the fitting function (24). The last coefficient is $c_{2}=-3\cdot 10^{-3}$.
\end{itemize}
\end{widetext}

\end{document}